# Visible luminescence from hydrogenated amorphous silicon modified by femtosecond laser radiation

Andrey V. Emelyanov[a], Andrey G. Kazanskii[a], Mark V. Khenkin[a], Pavel A. Forsh[a,b], Pavel K. Kashkarov[a,b], Mindaugas Gecevicius[c], Martynas Beresna[c], Peter G. Kazansky[c]

[a] M.V. Lomonosov Moscow State University, Physics Department, Moscow, Russia
[b] National Research Centre "Kurchatov Institute", Moscow, Russia
[c] Optoelectronics Research Centre, University of Southampton, Southampton, UK

Visible luminescence is observed from the composite of $SiO_2$ with embedded silicon nanocrystallites produced by femtosecond laser irradiation of hydrogenated amorphous silicon (a-Si:H) film in air. The photoluminescence originates from the defect states at the interface between silicon crystallites and $SiO_2$ matrix. The method could be used for fabrication of luminescent layers to increase energy conversion of a-Si:H solar cells.

Thin film technology based on hydrogenated amorphous silicon (a-Si:H) has been playing a significant role in the world production of photoelectric modules for several decades. However metastable structure of a-Si:H requires improvement to increase charge carrier mobility and reduce light-induced degradation. Nanocrystalline silicon (nc-Si:H) in comparison with a-Si:H attributes larger carrier mobility and higher stability, all of which are key elements for the design of thin film transistors or tandem amorphous-nanocrystalline solar cells. One of technologically attractive methods for a-Si:H crystallization is laser processing, which allows achieving localized structural modifications fast enough to minimize out-diffusion of hydrogen and the laser irradiation influence on a substrate [1]. Femtosecond laser writing in transparent materials has also recently attracted considerable interest for fabrication of integrated optics components, optical memory and photonic crystals [2]. Unlike excimer laser annealing, interaction of ultrafast near infrared laser radiation with amorphous silicon thin films involves nonlinear light absorption and non-equilibrium thermodynamics. Recently various experiments have been reported on femtosecond laser irradiation of a-Si:H films [3–5]. Most of the experiments were dedicated to the investigation of the effect of femtosecond laser processing on a-Si:H film crystallization, surface nano-structuring, optical absorption and photoelectric properties. In particular, ultrafast laser irradiation was demonstrated to produce simultaneous surface nano-structuring and crystallization of amorphous silicon film.

In this Letter we report observation of visible photoluminescence (PL) from a-Si:H films modified with femtosecond laser irradiation in air. The effect of high repetition rate femtosecond laser annealing on the luminescence properties of a-Si:H films is correlated with the change of their structure. We observed that both the PL intensity and the amount of oxygen incorporated into the silicon matrix increase with the laser fluence.

The a-Si:H films with a thickness of ~0.5 μm were deposited on silica glass substrates by plasma-enhanced chemical vapour deposition (PECVD) upon the decomposition of the mixture of silane ($SiH_4$) and argon (Ar) at substrate temperature of 250 $^0$C. The volume ratio between the gases in the reaction chamber corresponded to 25% $SiH_4$ and 75% Ar. Ultrafast laser treatment of a-Si:H films was carried out with Yb:KGW based femtosecond system (Pharos, Light Conversion Ltd.) that delivered pulses of 500 fs with repetition rate of 200 kHz and wavelength centered at 1030 nm. The laser beam was focused via the aspheric lens with a numerical aperture of 0.16. The focal plane was placed 80 μm above the sample surface in order to increase the laser irradiation area and avoid ablation. The Gaussian beam spot was circular with the beam diameter of about 15 μm on the film surface. The samples were irradiated by scanning with the translation speed of 5 mm/s. The scanning step was 2 μm. The average laser beam power was varied from 20 to 300 mW corresponding to the laser fluences from 30 to 480 mJ/cm$^2$. The laser processing was carried out in air.

Micro-Raman spectrometer (Horiba Jobin Yvon HR800) was used to examine the structure of irradiated samples. PL measurements were carried out at room temperature using Ar-laser as an excitation source (488 nm wavelength, 200 mW power) and 500 mm spectrometer equipped with a CCD camera for detection. PL spectra were corrected for the system response.

Figure 1 shows scanning electron microscope (SEM) images of the silicon film surfaces treated with different laser fluences. The SEM images of the samples revealed an abrupt change of the surface morphology after laser irradiation at the fluence of 250 mJ/cm$^2$. The additional atomic force microscope (AFM) imaging of the samples revealed surface nanospikes with heights of 30 – 40 nm formed after irradiation with the fluence values in the range of 30 – 200 mJ/cm$^2$. Spikes with the heights of 300 – 400 nm were created for the laser fluence values above 250 mJ/cm$^2$ and were responsible for the observed "black reflection" from the samples (Fig 1). Moreover, in these samples we were surprised to observe visible photoluminescence with the maximum at 675 nm, the intensity of which was increasing with the fluence (Fig. 2). The results indicate the existence of two different types of silicon film surface modification below and above threshold fluence of 250 mJ/cm$^2$.

To determine the origin of the observed photoluminescence we analyzed the chemical composition of the treated samples by X-Ray Photoelectron Spectroscopy (XPS) with PHI 5500 ESCA XPS spectrometer (Physical Electronics). The photoelectron emission was excited by Mg Kα radiation (hv = 1253.6 eV) with the power of 330 W. The diameter of the analyzed area ranged from 0.6 to 1.1 mm. The layer with a natural surface oxidation of about 4 nm thickness was removed from the top of the samples by Ar ion etching. The XPS spectra reveal oxidation by a shift toward higher binding energies. The binding energy of 2p electrons in bulk crystalline silicon is reported to be around 99.5 eV, whereas Si 2p electrons in $SiO_2$ have binding energies about 104 eV. Only silicon atoms were detected in the untreated samples, in contrast $SiO_2$ content increased dramatically after treatment with the fluence above 250 mJ/cm$^2$ (Fig. 3). The XPS spectra analysis of the treated samples revealed that about 90 % of Si atoms were oxidized after irradiation. This evidence of oxidation correlates with the observed abrupt change of the surface morphology of the samples. Additional ion etching experiment demonstrated that $SiO_2$ content remains constant at least within the surface layer of 50 nm thickness. The appearance and increase of both the PL intensity and XPS $SiO_2$ signal with the increase of laser fluence indicate that the photoluminescence originates from the oxidized silicon in the irradiated samples.

Surface oxidation of monocrystalline silicon wafer and visible photoluminescence upon femtosecond laser irradiation in air has been reported in [6, 7]. According to these results the laser irradiation melts the crystalline silicon surface, resulting in diffusion of oxygen into silicon and enhancing oxidation, due to six orders of magnitude increase of the oxygen diffusion coefficient in the liquid phase of silicon [8]. The increased surface area of the films subjected to high fluence irradiation can also contribute to the measured increase of SiO$_2$ content. Wu et al. [6] observed visible photoluminescence with the peak wavelength varying between 540 nm and 630 nm. After annealing the samples at 1300 K temperature they observed another photoluminescence band at lower energies. On the contrary, Chen et al. [7] observed two luminescence bands (near 600 nm and 680 nm) even without annealing. This discrepancy could be explained by the difference in repetition rates of the laser systems used in the experiments. Higher repetition rate could lead to heat accumulation, which would partially anneal the sample. In both cases, high (green/orange) and low (red) energy bands were respectively attributed to "defect luminescence" originating from the oxygen related defects at the interface between Si nanocrystals and SiO$_2$ matrix and "quantum confinement luminescence" attributed to the recombination of confined excitons in silicon nanoclusters. It is worth noting that the conclusion about the "quantum confinement luminescence" in [6] was based on the PL spectrum shift to lower energies upon thermal annealing of the laser treated crystalline silicon samples. According to [9, 10] to get such mechanism of luminescence with the peak wavelength around 700 nm silicon crystallites with the average sizes of 3 – 4 nm should be created after annealing of laser treated crystalline silicon samples.

Unlike [6, 7] where crystalline silicon wafers were used we could get additional information about the structure of our films, which exhibited visible photoluminescence, by means of Raman spectroscopy. The results are shown on Fig. 4. The Raman spectra for the untreated samples demonstrated spectrum typical for a-Si:H with the maximum near 480 cm$^{-1}$. While the Raman spectrum of high fluence treated a-Si:H film consisted of broad band at lower frequencies, that is apparently attributed to $SiO_2$ [11] and narrow silicon "crystalline" line with the peak position at 519.5 cm$^{-1}$ shifted by 1 cm$^{-1}$ with respect to that for monocrystalline silicon (520.5 cm$^{-1}$).

The Raman peak position shift $(\Delta\omega)$ due to the confinement effect is described by a confinement model [12, 13]

$$\Delta\omega = \omega(L) - \omega_0 = -A(a/L)^\gamma, \tag{1}$$

where $\omega(L)$ is the frequency of the Raman phonon in a Si nanocrystal with the size of $L$, $\omega_0$ is the frequency of the optical phonon at the zone center, and $a$ is the lattice constant of Si. The parameters $A$ and $\gamma$ are used to describe the vibrational confinement due to the finite size of a nanocrystal. We used in our estimates the parameters *A = 47.41* cm$^{-1}$ and *γ = 1.44*, obtained in [12] for Si spheres. We estimated the size of Si crystallites of about 8 nm according to the quantum confinement shift in Raman spectrum (1). According to data presented in [9] such crystallites should give luminescence at about 900 nm and quantum yield of less than 0.01. In our experiments, for a-Si:H films treated with the high fluence above 250 mJ/cm$^2$ the peak of PL spectra is near 675 nm and does not change its position with the laser fluence (Fig. 2). Thus we exclude the possibility of "quantum confinement luminescence", and nanocrystals detected by Raman spectroscopy could not be responsible for the observed photoluminescence (Fig. 2), which is most likely produced by the defect states, in particular non-bridging oxygen hole centers (NBOHC), at the interface between Si nanocrystals and $SiO_2$ matrix [14, 15].

Larger crystallites in the studied a-Si:H samples compared to those in treated crystalline silicon wafers [6, 7] could be explained by higher temperatures induced by the laser irradiation in the thin films of amorphous silicon deposited on fused silica substrates. Low thermal conductivity of the silica substrate leads to slower temperature decrease in the film compared to the surface of a crystalline silicon wafer.

In conclusion, we observed visible photoluminescence with the peak at 675 nm from the composite of Si nanocrystals and $SiO_2$, created by femtosecond laser irradiation of a-Si:H films in air. The luminescence intensity increases with the laser fluence used for film treatment. The photoluminescence is most likely associated with defects at the interface between Si nanocrystals and $SiO_2$ matrix. Demonstrated ultrafast processing offers the possibility of precisely localized $SiO_2$ formation in a-Si:H based solar cells. The observed red luminescence allows increasing the photosensitivity of solar cells to higher photon energies thanks to so called luminescence down shifter effect [16] that will result in several percent relative enhancement of the energy conversion efficiency.

**Acknowledgements**

The authors are grateful to Dr O. Konkov for a-Si:H films preparation, A. Khomich for Raman measurements and Dr. E. Skrileva for XPS measurements. This work was supported by the Ministry of Education and Science of the Russian Federation (contract No 16.513.11.3084), the Russian Foundation for basic Research (project 12-02-00751-a) and the project FEMTOPRINT, financed by the European Commission Factories of the Future program (FP7/NMP/Project No 260103).

**References**

1. P. H. Liang, C. J. Fang, D. S. Jiang, P. Wagner, and L. Ley, Appl. Phys. A **26**, 39 (1981).

2. G. J. Lee, J. Park, E. K. Kim, Y. P. Lee, K.M. Kim, H. Cheong, C.S. Yoon, Y-D. Son, and J. Jang, Optics Express **13,** 6445 (2005).

3. J. Shieh, Z. Chen, B. Dai, Y. Wang, A. Zaitsev, and C. Pan, Appl. Phys. Lett. **85**, 1232 (2004).

4. T. Y. Choi, D. J. Hwang, and C. P. Grigoropoulos, Opt. Eng. **42**(11), 3383 (2003).

5. A.V. Emelyanov, A.G. Kazanskii, P.K. Kashkarov, O.I. Konkov, E.I. Terukov, P.A. Forsh, M.V. Khenkin, A.V. Kukin, M. Beresna, and P. Kazansky, J. Semicond. **46**, 749 (2012).

6. C. Wu, C. H. Crouch, L. Zhao, and E. Mazur, Appl. Phys. Lett. **81**, 1999 (2002).

7. T. Chen, J. Si, X. Hou, S. Kanehira, K. Miura, and K. Hirao, J. Appl. Phys. **110**, 073106 (2011).

8. K. Hoh, H. Koyama, K. Uda, and Y. Miura, Jpn. J. Appl. Phys. **19**, 375 (1980).

9. G.Ledox, J.Gong, F.Huisken, O.Guillois, and C.Reinaud, Appl. Phys. Lett. **80**, 4834 (2002).

10. S. Takeoka, M. Fujii, and S. Hayashi, Phys. Rev. B **62**, 16820 (2000).

11. Z. Ma, X. Liao, G. Kong, and J. Chu, Sci. China, Ser. A Math. **43**, 414 (2000).

12. J. Zi, H. Buscher, C. Falter, W. Ludwig, K. Zhang, and X. Xie, Appl. Phys. Lett. **69**, 200 (1996).

13. G. Viera, S. Huet, and L. Boufendi, J. Appl. Phys. **90**, 4175 (2001).

14. W.J. Reichman, J.W. Chan, C.W. Smelser, S.J. Mihailov, and D.M. Krol, J. Opt. Soc. Am. B **24**, 1627 (2007).

15. S. Goderfroo, M. Hayne, M. Jivanescu, A. Stesnabs, M. Zacharias, O. I. Lebedev,

G. Van Tendeloo, and V. V. Moshalkov, Nature Nanotechnology **3**, 174 (2008).

16. F. Sgrignuoli, G. Paternoster, A. Marconi, P. Ingenhoven, A. Anopchenko, G. Pucker, and L. Pavesi, J. Appl. Phys. **111**, 034303-1 (2012).

**Figure captions**

**Figure 1:** SEM images of pristine a-Si:H film (top left) and irradiated with two different fluences (bottom left and right). White light reflection from the surface of a-Si:H thin film (top right). The reflection from the part of the sample irradiated at fluences more than 250 mJ/cm$^2$ is significantly reduced.

**Figure 2:** PL spectra of a-Si:H films treated by femtosecond laser radiation with laser fluence values of 260, 360 and 460 mJ/cm$^2$.

**Figure 3:** The part of the measured XPS spectrum corresponding to the Si 2p orbitals of untreated a-Si:H film (dashed line) and film treated with laser fluence of 260 mJ/cm$^2$ (solid line).

**Figure 4:** Raman spectra of untreated a-Si:H film (bottom line) and film treated with the laser fluence of 260 mJ/cm$^2$ (top line).

**Figures**

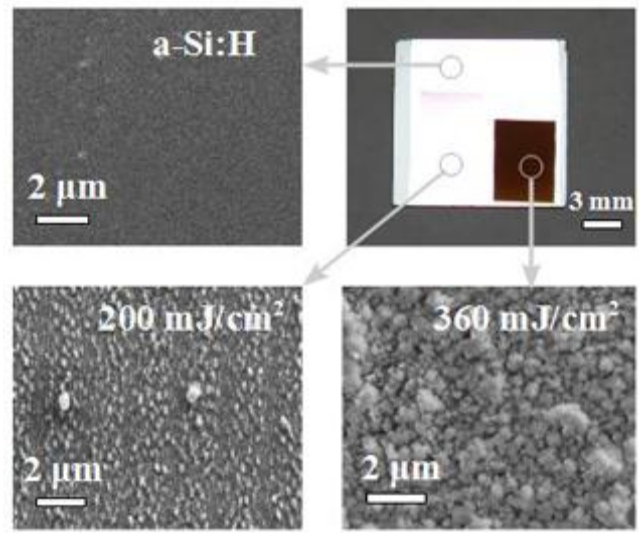

**Figure1.** Andrey V. Emelyanov et al.

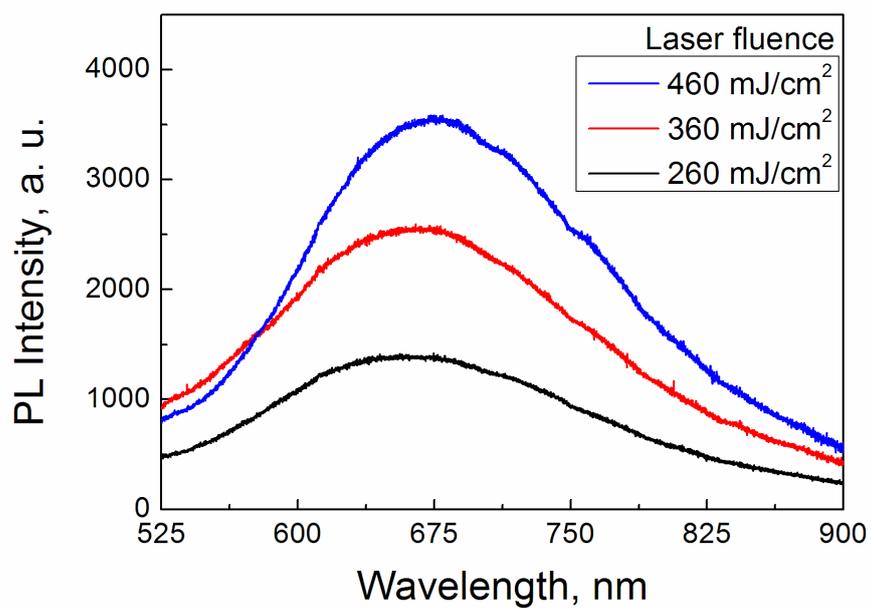

**Figure 2.** Andrey V. Emelyanov et al.

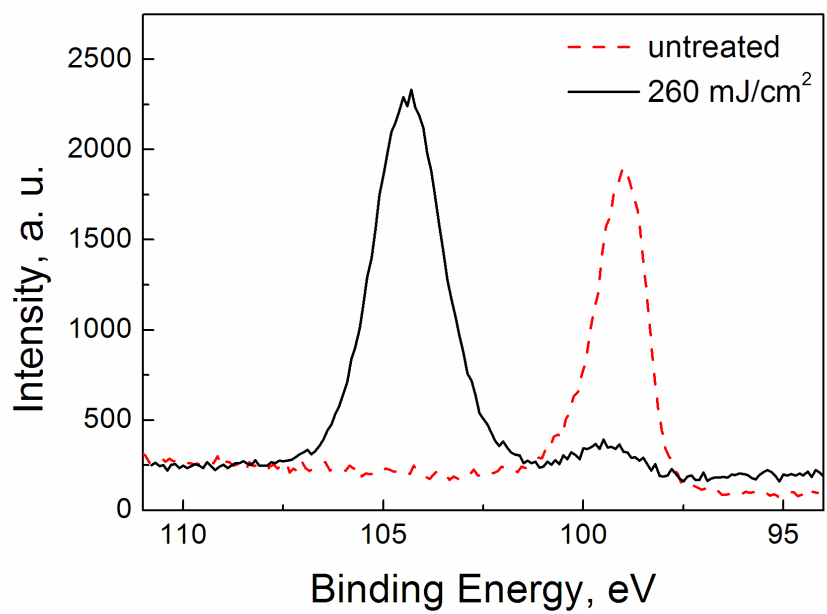

**Figure 3.** Andrey V. Emelyanov et al.

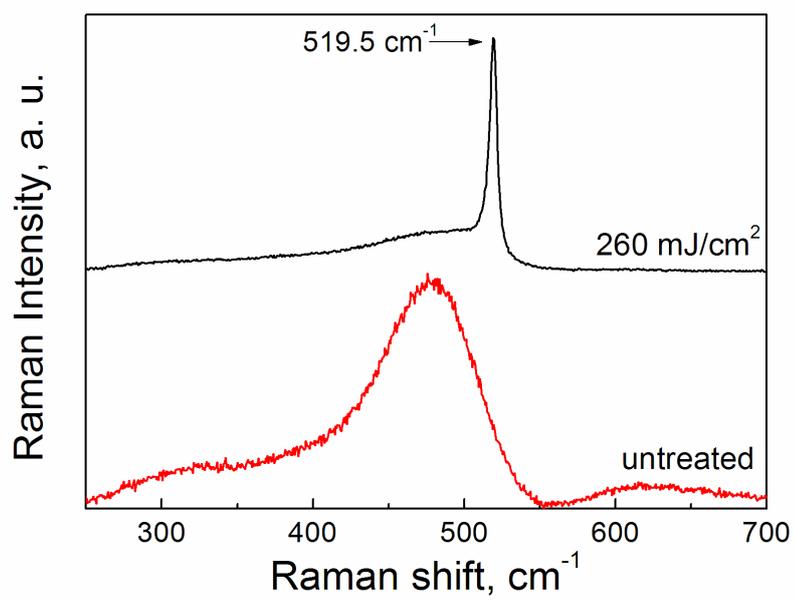

**Figure 4**. Andrey V. Emelyanov et al.